\documentclass[12pt]{article}

\usepackage{times}
\usepackage{graphicx}
\usepackage{scicite}
\usepackage{xcolor}

\usepackage{amsmath}
\usepackage{amsfonts}
\usepackage{physics}

\newcommand{\xx}{\mathbf{x}}
\newcommand{\kk}{\mathbf{k}}

\newenvironment{sciabstract}{%
\begin{quote} \bf}
{\end{quote}}

\title{Spontaneous symmetry breaking in diffraction}

\author
{J. Abad-Arredondo,$^{1}$ Z. Geng,$^{2}$ G. Keijsers, $^{2}$ F. Bijloo, $^{2}$ F. J. García-Vidal,
$^{1,3}$ \\ A. I. Fernández-Domínguez, $^{1\ast}$ S. R. K. Rodriguez $^{2}\dagger$\\
\\
\normalsize{$^{1}$Departamento de Física Teórica de la Materia Condensada and Condensed Matter Physics Center}\\
\normalsize{(IFIMAC), Universidad Autónoma de Madrid, E28049 Madrid, Spain}\\
\normalsize{$^{2}$Center for Nanophotonics, AMOLF,}\\
\normalsize{Science Park 104, 1098 XG Amsterdam, The Netherlands}\\
\normalsize{$^{3}$Donostia International Physics Center (DIPC), Donostia/San Sebastian 20018, Spain}\\
\\
\normalsize{$^\ast$ E-mail:  a.fernandez-dominguez@uam.es}\\
\normalsize{$\dagger$ E-mail: s.rodriguez@amolf.nl}
}

\date{}

\begin{document}


\baselineskip24pt


\maketitle

\normalsize{\textbf{One sentence summary:} Waves in a nonlinear grating can spontaneously acquire momentum and diffract at angles forbidden by Bloch’s theorem.  }



\begin{sciabstract}
The connection between symmetries and conservation laws is a cornerstone of physics. It underlies Bloch's theorem, which explains wave phenomena in all linear periodic systems. Here we demonstrate that, in a nonlinear grating with memory, diffracted waves can spontaneously acquire momentum parallel to the lattice vector in quantities unconstrained by the grating period. In this breakdown of Bloch’s theorem, which we also evidence in solutions to nonlinear Maxwell’s equations, wave amplitudes no longer respect the discrete translation symmetry of the grating. Our findings reveal a rich phenomenology for waves in nonlinear periodic systems,  and point to numerous opportunities for nonlinear lattices with broken symmetry in the context of imaging, sensing, and information processing in general.
\end{sciabstract}

According to Bloch's theorem (BT), wave amplitudes in a periodic potential must have the same periodicity as the potential itself~\cite{Bloch1929}.  This basic property of waves stems from the relation between symmetries and conserved quantities, first identified by Emmy Noether~\cite{Noether1918}. Essentially, BT is due to the discrete translation symmetry of the system and the corresponding conservation of the wavevector component parallel to the lattice vector, $k_{||}$. It may seem obvious that steady-state wave amplitudes must have the same symmetry as their confining potential. However, nature provides many examples of states with lower symmetry than their confining potentials. In fact, studies of spontaneous symmetry breaking (SSB) have shaped physics for decades~\cite{Anderson}.  For instance, the laser, Bose-Einstein condensation, superfluidity, superconductivity, the Josephson effect, and the Higgs boson, all emerge when a rotational U(1) symmetry is broken~\cite{Endres12, Navon15, Wezel19}. In addition, in atomic~\cite{Zibold10} and optical~\cite{Malomed13, Hamel15, Cao17, Garbin20, Xu21, Garbin22, Krasnok22, Hill23} systems, a mirror symmetry can spontaneously break and  localized states with quantum entanglement can emerge~\cite{Zoller03, Casteels17SSB}. In periodic systems, symmetry broken states have been theoretically analyzed~\cite{Kalozoumis}, but their spontaneous emergence and the concomitant breakdown of BT have never been observed.

Here we demonstrate the breakdown of BT triggered by SSB in diffraction. We measure light scattering from a nonlinear grating with memory  and, at sufficiently high intensities, we observe a cascade of dynamical effects in diffraction. These include spontaneous symmetry-breaking and symmetry-restoration transitions, as well as limit cycles and signatures of chaotic dynamics. Through numerical and analytical calculations at the level of Maxwell's equations, we explain our observations and the breakdown of BT. Our theoretical approach extends the use of linear stability analysis methods to extended arbitrary photonic structures, and illustrates how their refractive index fluctuation spectrum governs emergent phenomena in these structures.\\

\begin{figure}[b]
\centering
\includegraphics[width=1\linewidth]{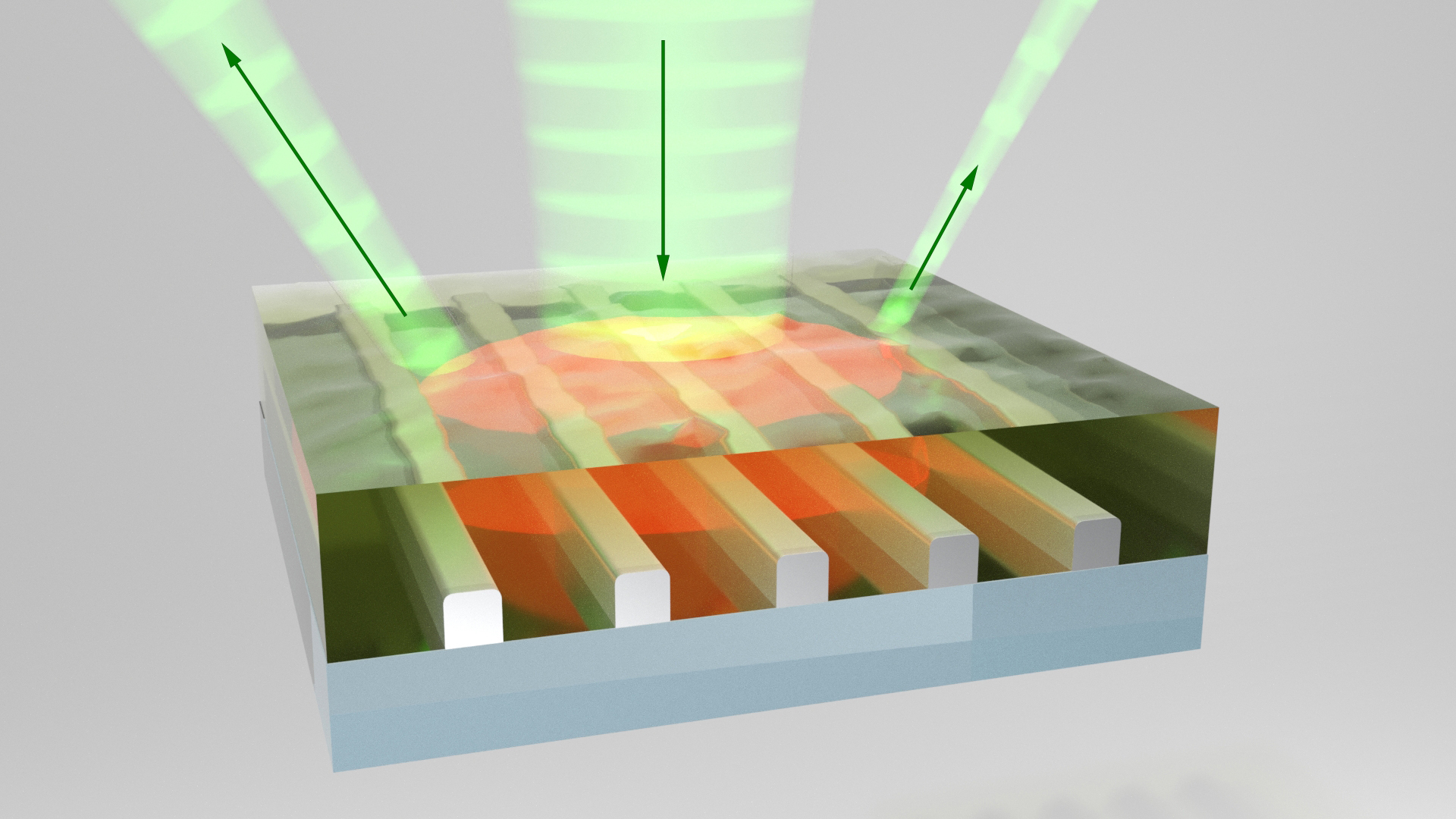}
\caption{\textbf{Spontaneous symmetry breaking in diffraction.}  An unblazed metallic grating coated with oil is illuminated by a laser at normal incidence. Due to the thermo-optical nonlinearity of the  oil, the symmetry of the diffraction pattern can spontaneously break implying a breakdown of Bloch's theorem.}
\label{fig:1}
\end{figure}

\noindent\normalsize{\textbf{Experimental observation of spontaneous symmetry breaking in diffraction}}\\
\noindent Figure 1 illustrates the system under study: a one-dimensional grating coated with cinnamon oil. The grating stands on a glass substrate, and comprises aluminum wires with 90 nm width, 70 nm height, and 366 nm lattice constant. A 532 nm continuous wave laser impinges perpendicular to the periodicity plane.  The laser wavelength is close to a grating resonance, as shown in Supplementary fig. S2. Part of the laser light is absorbed by the oil, and then dissipated as heat. The resultant temperature rise causes the oil to expand, and its density and refractive index to decrease. This process, a thermo-optical nonlinearity, corresponds to an intensity-dependent refractive index. Unlike in the Kerr effect where the refractive index changes instantaneously, here the refractive index change is delayed by the finite thermal relaxation time of the oil. The non-instantaneous response effectively gives memory to the system~\cite{Geng_PRL2020,Peters20}.

Figure 2(A) shows the sample's transmittance  when the laser power is temporally modulated. The transmittance depends on the power and the direction of the power scan. This irreversibility, or hysteresis, is sometimes taken as an indication of bistability~\cite{Wurtz06}: two stable states at a single driving condition. However, a system can display hysteresis without bistability~\cite{Strogatz97,Rodriguez17}. A stronger evidence of bistability is the abrupt jump in transmittance at 2.39 s, which signals a transition between states.  In Supplementary fig. S4 we plot the transmittance versus input power, evidencing a wide bistability region. In addition, a zoomed-in view of the transmittance shows an undershoot after the jump. The half width at half maximum of the undershoot, 60 $\mu$s, is indicative of the thermal relaxation time, $\tau$~\cite{Geng_PRL2020,Peters20}. We also recorded images of the grating’s reflection as a function of power. Figures 2(B,C,D) show three images taken 50-70 ms after the jump.  The white disk and rings around the center of all images are due to the direct laser reflection. The dots enclosed by dashed white circles are due to +1 and -1 diffraction orders (see Supplementary fig. S3 and discussion around it for details).

\begin{figure}[t]
\centering
\includegraphics[width=0.5\linewidth]{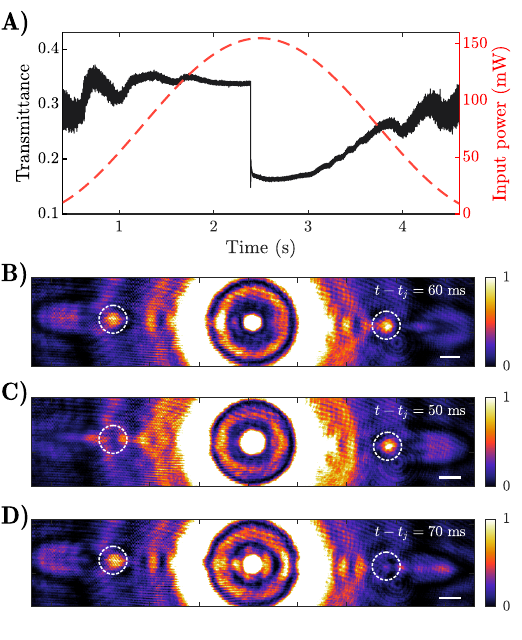}
\caption{\textbf{Optical bistability and spontaneous symmetry breaking in diffraction.} In the system illustrated in Fig. 1, we modulate the incident laser power (red dashed curve) and measure the transmittance (black curve). The jump around $t_j=2.4$ s and the existence of two states with different  transmittance at a single input power demonstrate optical bistability. (B, C, D) Reflection images taken shortly after the jump time $t_j$, as indicated in the top right corner. The bright dots enclosed by the dashed white circles correspond to the  +1 and -1 diffracted orders. The dramatic change in relative intensities  of the  +1 and -1 diffracted orders indicates SSB. Scale bar indicates 3 $\mu$m.}
\label{fig:2}
\end{figure}

\begin{figure}[t]
\centering
\includegraphics[width=0.5\linewidth]{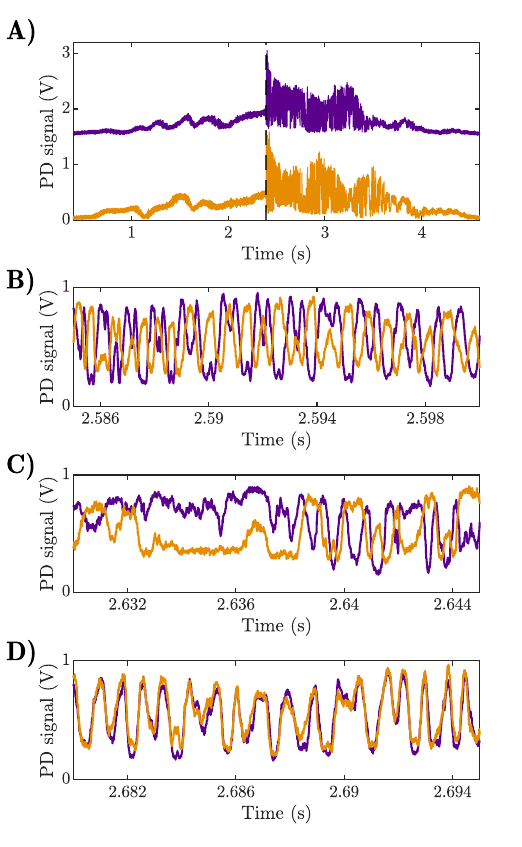}
\caption{\textbf{Limit cycles and chaos in diffraction.} (a) Diffracted intensities by the -1 (purple) and +1 (orange) diffraction orders, respectively, indicated by the dashed circles in Fig. 2. The purple curve is displaced vertically for clarity. (B, C, D) Zoom into three intervals of 15 ms  in (A). The diffracted intensities oscillate out of phase in (B), in phase in (D), and randomly in (C).}
\label{fig:3}
\end{figure}

The +1 and -1 diffracted intensities are similar in Fig. 2(B), but different in Fig. 2(C) and Fig. 2(D) which were taken 10 ms before and 10 ms after Fig. 2(B), respectively. Similar changes in relative intensities occurred suddenly in various scans, and without noticeable tendency for the +1 order or the -1 order to become brighter more often. Since the grating is symmetric (unblazed) and the laser impinges perpendicular to the periodicity plane, we interpret these intensity changes as SSB in diffraction.  To investigate this effect with greater temporal resolution,  we constructed a setup for isolating the two diffracted intensities (inside the dashed circles) from the background reflection and sending them to photodetectors. A sketch of the setup is in Supplementary fig. S5, and measurement details are in Methods. The results are shown in (Fig. 3), for the same modulation of the input power as in Fig. 2(A). Purple and orange curves correspond to the -1 and +1 diffracted intensities, respectively.

Figure 3(A) displays rich dynamics of the diffracted intensities immediately after the jump, indicated by the vertical dashed black line.  Figures 3(B,C,D) zoom into three representative time windows, each 15 ms long. Figures 3(B) and \ref{fig:3}(D) show out-of-phase and in-phase oscillations, respectively, of the +1 and -1 diffraction orders. Such self-sustained oscillations, known as limit cycles~\cite{Strogatz18}, are here observed for the first time in diffraction. In between these two limit cycles, we observe  a window of chaotic, uncorrelated dynamics as Fig. 3(C) shows.

The phase of the limit cycle is chosen spontaneously, similar to the spontaneous choice of the phase of a scalar field at a  rotational U(1) symmetry breaking transition~\cite{Wezel19}. For this reason, limit cycles have drawn interest as manifestations of ‘time crystals’ --- self-organized periodic states in time emerging through SSB~\cite{Shapere12,Yao20,Marconi20,Kessler22,oded23}. Interestingly, the in-phase oscillations in Fig.\ref{fig:3}(d) respect the spatial symmetry of the system, but the out-of-phase oscillations in Fig.\ref{fig:3}(B) do not. This possibility, namely for spatial symmetry to be broken or not in a time crystalline phase, was recently analyzed in a model of coupled cavities~\cite{Lledo2020}. Here, we evidence this phenomenon by breaking and restoring spatial symmetry as the driving power increases.

We now address the important question of whether the spatial symmetry of our system is indeed broken spontaneously by fluctuations, or explicitly by an unaccounted bias. The excellent overlap of the two trajectories in Fig. 3(D) demonstrates the absence of detectable bias in our experiments. Such an excellent  overlap only occurs under symmetric driving conditions. We verified that, under explicit symmetry breaking (deliberate setup misalignment), the oscillations in the synchronized state no longer overlap. Thus, the fidelity of the symmetry restoration transition between Fig. 3(C) and Fig. 3(D) attests to the spontaneous character of  symmetry breaking transitions in the same system.\\

\begin{figure}[t]
\centering
\includegraphics[width=\linewidth]{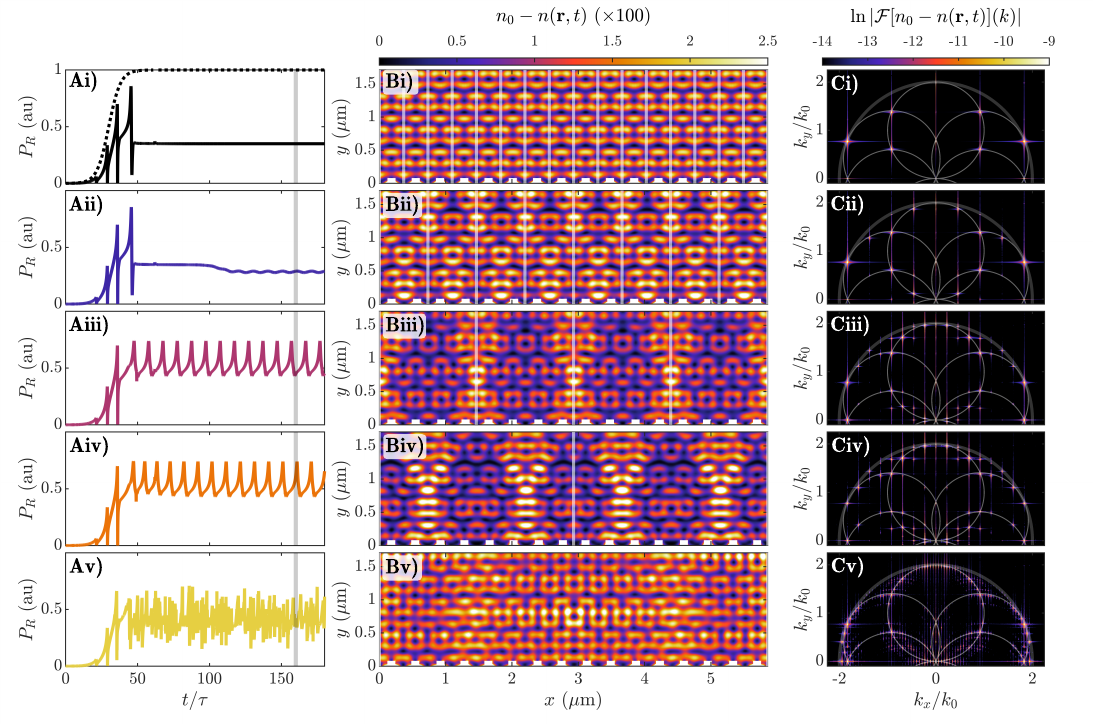}
\caption{\textbf{Breakdown of Bloch's theorem in numerical solutions of Maxwell's Equations.} Roman numeral labels indicate the size of the system supercell: 1, 2, 4, 8, and 16 grating periods, respectively. (A) Reflected power as a function of time. Dotted line on panel a(i) shows the input power profile. (B) Spatial refractive index profile at $t=160\tau$, see vertical grey line in panels (A). Note that the maximum index modification is 0.025, which is much smaller than $n_0=1.59$. Vertical white lines indicate the supercell size. (C) Fourier transform (in natural log scale) of the refractive index maps in panels (B).}
\label{fig:4}
\end{figure}

\noindent \normalsize{\textbf{Spontaneous discrete translation symmetry breaking in  Maxwell's equations}}\\
\noindent Next we discuss two complementary approaches to explain our experimental observations based on Maxwell's equations. In both of them, the oil layer is characterized by an intensity-dependent refractive index with memory of the form
\begin{equation}
n({\bf r},t)=n_0-\gamma\int_{-\infty}^t ds K(t-s) |{\bf E}({\bf r},s)|^2.
\label{eq:1}
\end{equation}
$n_0$ is the linear refractive index, $\gamma$ is the thermo-optical nonlinearity strength, ${\bf E}({\bf r},t)$ is the electric field, and $K(t)=e^{-t/\tau}/\tau$ is a memory kernel as used for single-mode oil-filled cavities~\cite{Geng_PRL2020,Peters20}. $\tau$ is the thermal relaxation time of the oil introduced above, which is also the memory time of the system. We took all parameter values from experiments, and  validated our model by reproducing the linear  spectrum (see figs. S1 and S2).\\

Our first approach involves full-wave simulations of nonlinear electromagnetic scattering under plane wave illumination. We solved for the electric fields in frequency domain as ${\bf E}({\bf r},\omega,t)$ ($t\sim\tau \gg 2\pi/\omega$). We avoided time-domain simulations by decoupling thermal and optical effects, which is justified for $\tau \gg \Gamma^{-1}$ with $\Gamma$ the optical dissipation rate; $\tau / \Gamma^{-1} \sim 10^9$ in our experiments. Crucially, we use mirror conditions on the lateral boundaries of the simulation domain. These are equivalent to Floquet periodic conditions for normal incidence and linear response but, unlike them, allow for wavevector components that do not comply with Bloch's theorem (BT) to emerge.

Figures 4A(i-v) show the calculated reflected power dynamics for simulations containing 1, 2, 4, 8 and 16 grating periods in the simulation domain, which acts as the system supercell.  In all cases, we ramped the incident power density as shown by the dashed curve in Fig. 4A(i). For short times ($t<40\tau$), the reflection undergoes fast oscillations akin to those observed experimentally. The oscillations are associated with the intensity-dependent resonant excitation of guided modes in the oil, as shown in Supplementary Material. The supercell size has negligible influence on this effect.

After the maximum incident power is reached and the system is free to evolve, a complex behavior emerges depending on the supercell size. This includes regimes of self-sustained oscillations and random dynamics in reflection like those observed experimentally. Figures 4B(i-v) display  $n_0 - n({\bf r},t)$ at the time indicated by the vertical grey line in panels A(i-v) and across 16 grating periods. Recall from Eq. 1 that
 $n({\bf r},t)$  is proportional to the intensity $|{\bf E}({\bf r},\omega,t)|^2$.  White rectangles indicate the metallic wires, and vertical white lines indicate the supercell size. In Fig. 4B(i), where supercell and grating period are equal, the periodicity of $n({\bf r},t)$ matches that of the grating as expected from BT.  In contrast, for the larger supercells in Figs. 4B(ii-v), $n({\bf r},t)$  has a different periodicity than the grating and BT no longer holds. Remarkably, the  discrete translation symmetry of the linear system is broken. Supplementary fig. S6 demonstrates that the electromagnetic fields, which follow the refractive index maps in Figs. 4B, indeed do not comply with BT.

Refractive index  maps in Figs. 4B(ii-v) are time dependent, and self-sustained oscillations with different period emerge depending on the supercell size. Oscillations are smooth and take time to develop after the power ramp in Fig. 4A(ii), while they are sharp and emerge without delay after the power ramp in Fig. 4A(iii,iv). In Fig. 4A(v) we observe fast random dynamics,  a fingerprint of chaos. The entire phenomenology suggests that the system is driven deeper into the nonlinear regime as the simulation domain increases. Since the incident power density is kept constant, the nonlinear threshold is effectively reduced with increasing supercell size.

In Figs. 4C(i-v) we present the Fourier transform, $\mathcal{F}[\cdot]$, of the refractive index maps in Figs. 4B(i-v). Wavevectors are normalized to the homogeneous medium reference, $k_0=n_0 \omega/c$.  Most reciprocal space contributions fall on the circles of radius $k_0$, indicated  by thin white curves. The thicker white circle of radius $2k_0$ encloses all the wavevectors in $n({\bf r},t)$ ($\propto|{\bf E}|^2$) that can be excited by propagating plane waves. The peaks in Fig. 4C(i) all lie at $k_{||}=$ $k_x=m 2\pi/a$, with $m=0,\pm1,\pm2$, satisfiying BT. In contrast, for larger supercells, wavevector contributions violating BT emerge. In Supplementary fig. S6 we show that these wavevector components are responsible for the reflection oscillations in  Fig. 4A(ii-iv). We thereby establish a connection between the breakdown of BT and the emergence of limit cycles in our system. Deeper into the nonlinear regime, Fig. 4C(v) shows a nearly homogeneous  reciprocal-space peak density  along the thin white circles (see Supplementary fig. S9, and discussion around it for further insights on this result). This broadening of the angular spectrum suggests that the random dynamics in Fig. 4A(v) are indeed due to chaos. Our results above show that mirror boundary conditions enable local translation SSB and self-sustained oscillations to emerge. However, the global inversion SSB responsible for the sudden change in the  $+1$ and $-1$ diffracted intensities cannot occur under those boundary conditions.\\

\noindent \normalsize{\textbf{Asymmetric diffraction from a symmetric system}}\\
\noindent To explain the emergence of asymmetric diffraction from a symmetric and symmetrically-driven grating, we conceived a second approach using a Born scattering series to first order~\cite{Novotny12} and seeking a self-consistent solution to Maxwell's equations fed with Eq. 1. We treat the nonlinearity perturbatively, which is justified because the maximum index modulation in Fig. 4B(i-v) is ${0.025\ll1.59=n_0}$. After a linearization technique described in Supplementary Section S5, we find that fluctuations to the refractive index map in reciprocal space, $\delta n({\bf k},t)=\mathcal{F}[n({\bf r},t)-n_0]$, satisfy
\begin{equation}
\tau\frac{d}{dt}\delta n({\bf k},t)+\delta n({\bf k},t)=
\frac{2\gamma E_0^2}{n_0\chi}\sum_{\alpha\beta} M_{\alpha,\beta}({\bf k})\delta n({\bf k}+{\bf k}_\alpha-{\bf k}_\beta,t).
\label{eq:2}
\end{equation}
\noindent $\chi$ is the refractive index loss tangent (set according to experimental measurements), $E_0$ is the incident plane wave amplitude,  and $\alpha$, $\beta$ label the diffraction orders present in the linear solution.
The ratio $\gamma E_0^2 / \chi$ quantifies the balance between driving and dissipation, which determines the onset of the nonlinear regime. The matrix $M_{\alpha\beta}({\bf k})$ (analytical expression in Supplementary Material) describes the coupling between refractive index components of different wavevectors. It corresponds to the Jacobian matrix employed to study the stability of a fixed point of a dynamical system, and to the Bogoliubov matrix used to assess the excitation spectrum of a condensate. Importantly, $M_{\alpha\beta} ({\bf k})$ diverges for $\bf k$ at a distance $k_0$ from the wavevectors of the linear solution. This condition indicates which reciprocal-space components beyond BT can emerge in $n({\bf r},t)$, and coincides with the thin white circles in Figs. 4C(i-v).

Equation 2 is non-local in reciprocal space, resulting in a complex interplay between different refractive index fluctuations. For instance, high-wavevector components excited by the evanescent fields at the metal grating can serve as a seed for fluctuations of arbitrary wavevectors. If the system crosses the nonlinear threshold, refractive index fluctuations that do not respect BT can exponentially grow over time and govern $n({\bf r},t)$. The properties of these fluctuations are determined by the eigenvalue problem $\sum_{\alpha\beta} M_{\alpha\beta}({\bf k})\delta \tilde{n}_\lambda({\bf k}+{\bf k}_\alpha-{\bf k}_\beta,t)=\lambda\delta \tilde{n}_\lambda({\bf k},t)$ corresponding to Eq. 2, and their dynamics satisfy
\begin{equation}
\delta\tilde{n}_\lambda({\bf k},t)=\delta\tilde{n}_\lambda({\bf k},0)\exp{\left[\left(-1+\frac{2\gamma E_0^2}{n_0\chi}\lambda\right)\frac{t}{\tau}\right]}.          \label{eq:3}
\end{equation}
Defining a critical intensity $E_c^2=n_0\chi/2 \gamma {\rm Re}\{\lambda\}$,  a fluctuation is amplified or attenuated if $E_0>E_c$ or $E_0 <E_c$, respectively. Therefore, refractive index fluctuations that dominate the nonlinear dynamics are those whose eigenvalues have the largest positive real part. Equation 3 also reveals that oscillatory dynamics are governed by ${\rm Im}\{\lambda\}$, and, in agreement with our experiments, their period depends on $\tau$ and on the incident power.

\begin{figure}[t]
\centering
\includegraphics[width=0.8\linewidth]{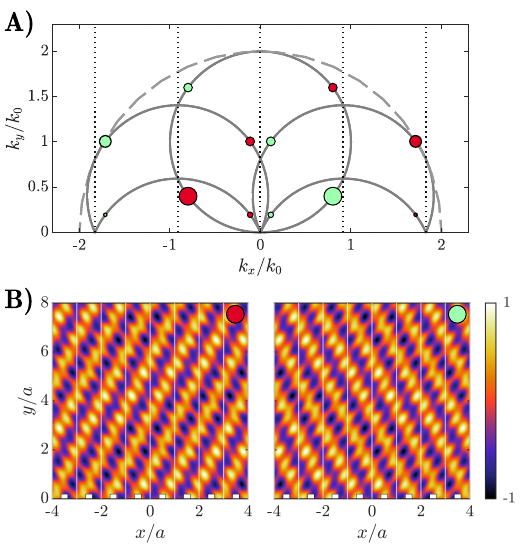}
\caption{\textbf{Linear stability analysis and global symmetry breaking.} (a) Reciprocal-space representation of the two (red and green) degenerate eigenfunctions, $\delta \tilde{n}_\lambda({\bf k},t)$ with largest ${\rm Re}\{\lambda\}$ for 64 grating periods. Dot sizes code the eigenfunction amplitudes. Vertical dotted lines indicate the wavevectors allowed by Bloch's theorem for the metallic grating. (b) Real space representation of the soliton-like refractive index fluctuations in panel (a).}
\label{fig:5}
\end{figure}

For the system in Fig. 4, we tackled the eigenvalue problem numerically (see supplementary Section S5.2 and fig. S7 therein.), and found that the maximum ${\rm Re}\{\lambda\}$ increases, and $E_c$ therefore decreases, as the supercell size increases. This behavior of the eigenvalue spectrum, shown in Supplementary fig. S8, explains the reduction in power needed to access the nonlinear regime as the supercell size grows (see previous section).   Our calculations also show that the maximum ${\rm Re}\{\lambda\}$ saturates to a constant value for supercells larger than 16 grating periods, indicating that our results are representative of the continuum limit. Figure 5(a) illustrates the two eigenfunctions $\delta \tilde{n}_\lambda({\bf k},t)$,  (red and green dots) associated with the dominant degenerate eigenvalue for a supercell spanning 64 grating periods. Since the eigenfunctions are extremely pointed around a discrete set of wavevectors, only the largest wavevector components (amplitude encoded by the dot size) are shown for clarity. The $k_x$ values allowed by BT are indicated by vertical dotted lines. Remarkably, the dominant wavevector contributions in our system deviate from those lines, thereby showcasing the breakdown of BT. The degenerate eigenfunctions are mirror images of each other, as shown in Fig. 5(b) which renders them in real space. Crucially, above the nonlinear threshold, any of these eigenfunctions can be excited by random fluctuations, resulting in SSB. We observe this phenomenon for all supercell sizes larger than one grating period, also in this model. We therefore infer that, in the continuum limit, our analytical approach elucidates the spontaneous emergence of an asymmetric radiation pattern from an unblazed and symmetrically driven grating. \\

\noindent \normalsize{\textbf{Conclusions and perspectives}}\\
\noindent To summarize, we have shown how the discrete translation symmetry of light in a nonlinear grating can spontaneously break, and restore, upon a continuous change in intensity. This phenomenon, implying a breakdown of Bloch's theorem, manifests through limit-cycle and chaotic dynamics in diffraction. We explained our findings through electromagnetic simulations and a linear stability analysis, which furthermore demonstrated how steady states with non-zero in-plane momentum can spontaneously emerge in perfectly periodic systems under normal plane-wave illumination. Our results open many opportunities for manipulating light without the constraints imposed by the symmetries of the system it interacts with. Taking advantage of recent advances in photonic materials and design maximizing light-matter interaction times~\cite{Khurgin23}, we foresee the implementation of stronger, faster and tunable SSB phenomena in space and/or time in different areas. On one hand, SSB offers an unprecedented dynamical control over scattered light momenta and wavefronts which is promising for spatial light modulators and super-resolution imaging~\cite{Forbes21}. On the other hand, SSB in spatially-extended systems like ours can be exploited for biosensing applications which may require integration with microfluidics and without the need of nanophotonic field confinement~\cite{Altug22}. Finally, by replacing our simple grating with more complex nanophotonic structures and illumination schemes, all-optical artificial neural networks for beyond von Neumann computing~\cite{Lin18} may be realized.\\

\vspace{8 mm}

\noindent \textbf{Methods}\\
\noindent \textbf{\textit{Sample and experiments}}\\
\noindent The aluminum grating is  $300\times 300\,\mu{\rm m}^2$. It was fabricated using standard electron-beam lithography and lift-off methods. On top of the grating we placed a drop of cinnamon oil. By pressing on the oil with a glass plinth of area $500 \times500\,\mu{\rm m}^2$, we set the thickness of the oil layer over the grating to 30 microns. We used piezoelectric actuators to control the position and orientation of the plinth, as well as of the grating. The actuators enabled us to align the plinth and grating parallel to each other, and  perpendicular to the optical axis.

We illuminated the grating-oil system with a single-mode continuous wave 532 nm laser. The laser impinged at the center of the grating, so that effects due to the finite size of the grating can be excluded. We modulated the laser intensity in time using a polarizing beam splitter and a half waveplate mounted on a motorized rotary stage. We used another half waveplate to ensure that the incident light polarization was parallel to the grating vector. For both optical excitation and collection, we used microscope objectives with 10x magnification and $NA = 0.25$ numerical aperture. The laser beam had a diameter of $4$~mm at the entrance of the objective. Since this is less than the $10$~mm aperture of the objective, the laser beam was loosely focused onto the grating plane. The transmitted intensity was measured by a photodetector, and the reflection was measured by either a camera or a pair of photodetectors. The camera in reflection was used to record images such as the ones in Figs. 2(b,c,d), while the pair of photodetectors was used to measure the diffracted intensities as shown in Fig. 3. On the optical path leading to each of the two photodetectors in reflection, we created two image planes where we placed pinholes. These pinholes were carefully positioned in order to isolate the diffracted intensities (signals insides the white dashed circles in Fig. 2) from the background. All photodetectors had a 50 MHz bandwidth, which is well above the thermal relaxation rate.\\

\noindent \textbf{\textit{Electromagnetics simulations}}\\
\noindent Full-wave electromagnetic simulations were performed in COMSOL Multiphysics. Applying translational invariance along grating lines, simulations were performed in 2D. The system is excited by a plane wave at normal incidence with a wavelength of $\lambda= 532$ nm, polarized along the grating vector. The nonlinear oil layer was set to 30 $\mu$m as in the experiments. The simulation domain was terminated vertically by a perfectly matched layer and scattering boundary conditions. Convergence studies were performed on the mesh size to ensure accurate results. To simulate the dynamics we used the COMSOL Livelink with Matlab to implement a first order finite differences scheme. Convergence studies were performed on the time stepping.\\

\noindent \textbf{\textit{Linear Stability Analysis}}\\
\noindent  self consistent solution to Maxwell's equations was found by using the Born scattering series to the first order. This allows to frame the problem entirely in terms of the refractive index change of the oil and the electromagnetic solution of the linear system (See details in SM). This effectively linearizes the nonlinear problem around the linear solution, instead of around the zero field solution. The problem is then cast as an eigenvalue problem. The momentum discretization mesh was generated using COMSOL Multiphysics built-in mesh building tools. The electric field solution for the linear problem was obtained using a coupled-quadrupole model adapted from  \cite{Swiecicki2017} to our geometry. Numerical diagonalization was performed using Matlab's built-in numerical diagonalization routines.

\bibliographystyle{Science}

\section*{Acknowledgments}
\noindent We thank Femius Koenderink for discussions.\\

\noindent \textbf{Funding}\\
\noindent This work is part of the research programme of the Netherlands Organisation for Scientific Research (NWO). S.R.K.R. acknowledges an ERC Starting Grant with project number 852694. J.A.-A., A.I.F.-D. and F.J.G.-V. acknowledge funding from the Spanish Ministry of Science, Innovation and Universities through Grants Nos. PID2021-126964OB-I00, PID2021-125894NB-I00, and TED2021-130552B-C21, as well as the European Union’s Horizon Programme through grant 101070700 (MIRAQLS).\\

\noindent \textbf{Author contributions} \normalsize \\
\noindent S.R.K.R. conceived the work. Z. G. and S.R.K.R. performed the experiments, with contributions from  G.K. and F.B. J.A.-A. performed the theoretical analysis, under the supervision of A.I.F.D., and F.J.G.-V. S.R.K.R., A.I.F.D., and J.A.-A. wrote the manuscript, with contributions from all authors. All authors discussed the results and the manuscript. \\

\noindent \textbf{Competing interests} \normalsize \\
\noindent The authors declare no competing interests.\\

\noindent \textbf{Data availability} \normalsize \\
\noindent Datasets generated during the current study are stored in a replication package within the AMOLF server. The replication package is available from the corresponding author on request. In addition, data for all figures in this manuscript will be uploaded to the Zenodo repository before publication. \\

\noindent \textbf{Code availability} \normalsize \\
\noindent Codes for data analysis and numerical calculations are part of the replication package mentioned above and will be uploaded to the Zenodo repository before publication. \\

\renewcommand{\thefigure}{S\arabic{figure}}
\setcounter{figure}{0}
\clearpage

\begin{center}
\title{\LARGE Spontaneous symmetry breaking in diffraction:\\ Supplemental Information}
\end{center}

\section{Characterization of the linear system}
\begin{figure}[h]
\centering
\includegraphics[width=\textwidth]{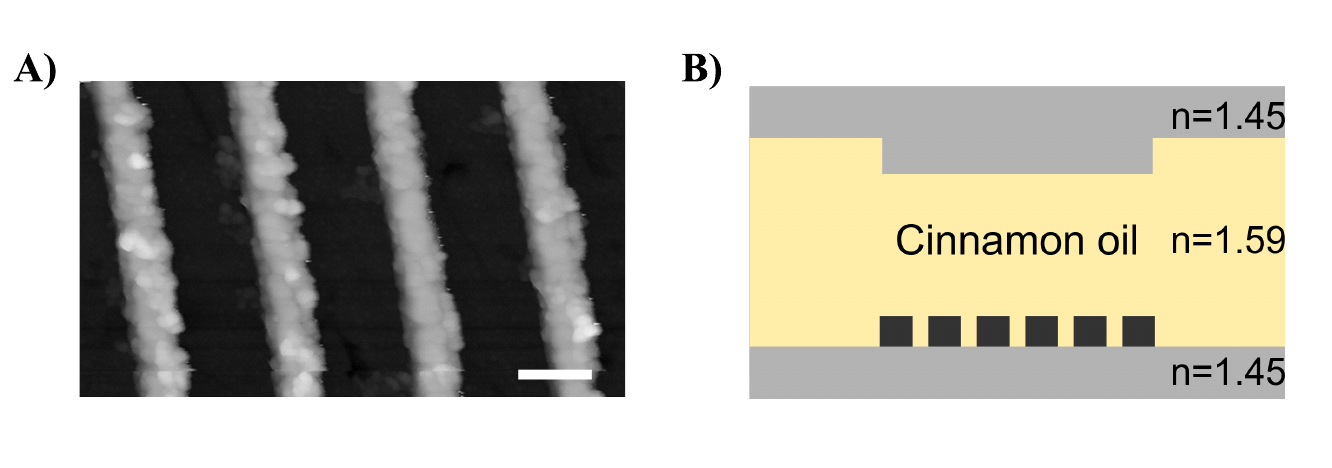}
\caption{\textbf{Sample.} A) Scanning electron micrograph of the aluminium grating studied in the main text. Scale bar is 200 nm. B) Side-view schematic of the the aluminum grating in A) coated with cinnamon oil. The height of the oil layer is controlled by pressing it with a glass plinth from above.}
\label{fig:S12}
\end{figure}

In this section we report the dimensions and linear optical properties of the experimental system. Using standard electron beam lithography and lift-off methods,  we  fabricated a grating of aluminium nanowires on a glass substrate. The wires have an approximately rectangular cross section, 90 nm wide and 70 nm tall. The grating period is 366 nm. A scanning electron micrograph of the grating is shown in fig.~\ref{fig:S12}(A).  For the optical experiments, we coated the grating with a layer of cinnamon oil. The height of the oil layer was controlled by pressing on it with a glass plinth, which was mounted on  a piezoelectric actuator. A side-view schematic of the sample is shown in fig.~\ref{fig:S12}(B).

In the linear regime, the glass substrates and oil layer are  each characterized by a constant refractive index $n_g=1.45$ and $n_0=1.59$, respectively. For the aluminum nanowires we use the refractive index in  Ref. \cite{Rakic1995}.

\subsection{Linear spectra: measurements and simulations}
\begin{figure}[h]
\centering
\includegraphics[width=0.9\textwidth]{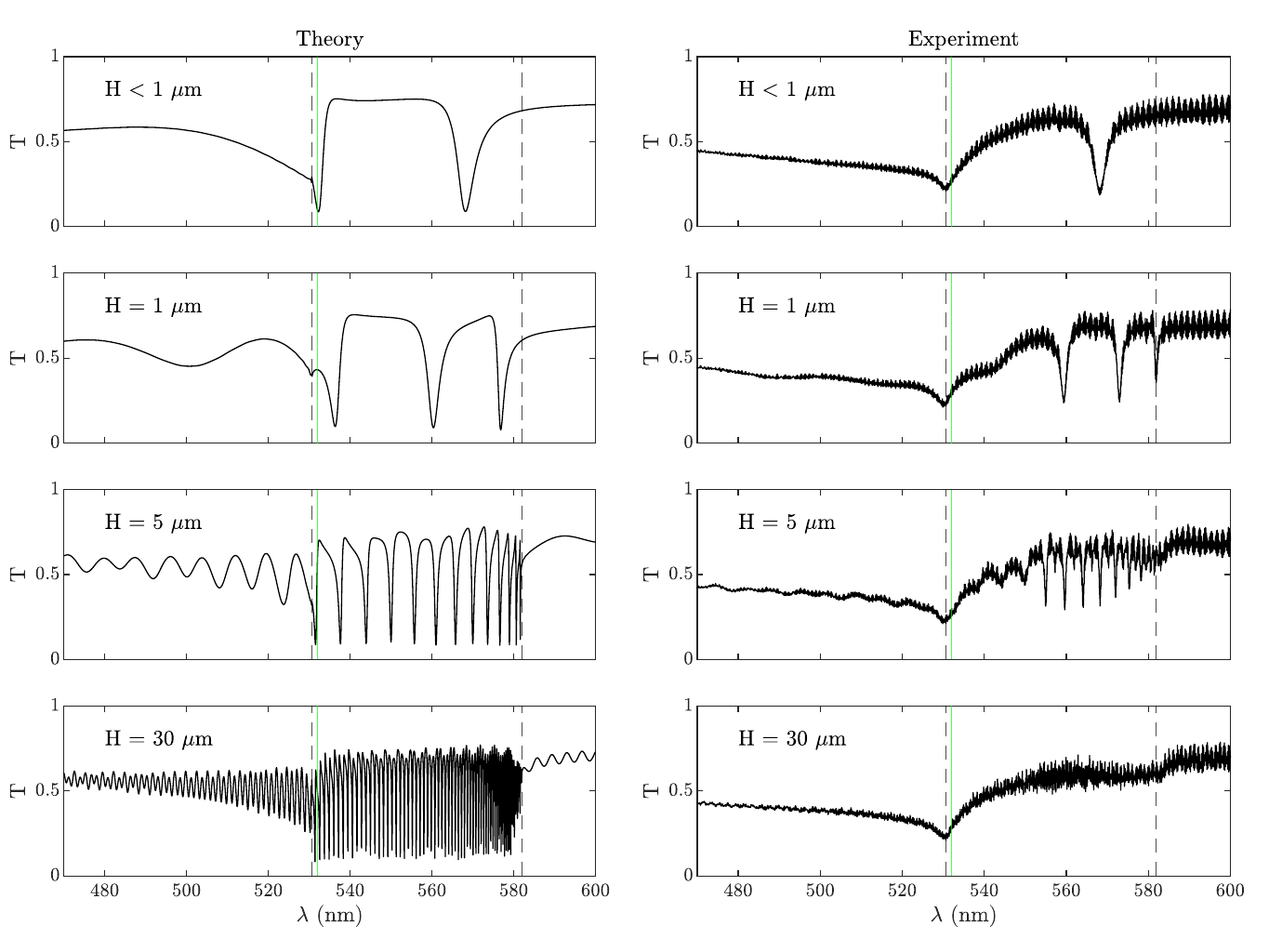}
\caption{\textbf{Linear transmittance spectra.} Simulated (left column) and experimentally measured (right column) linear transmittance spectra for different oil layer heights. Vertical dashed lines indicate the wavelengths at which the first diffraction order changes from radiating to evanescent in the glass ($\lambda= 531$ nm) and   oil ($\lambda=582$ nm). Vertical green line indicates the wavelength of the laser used for the experiments in the main text.}
\label{fig:S10}
\end{figure}
The period $a$ of the grating determines the reciprocal grating vector $k_a=2\pi/a$ that the system will exchange with any plane wave. For an incident plane wave of wavevector $\vec{k}=K\hat{x}-\sqrt{k^2-K^2}\hat{y}$, the diffracted orders will exchange $N$ reciprocal lattice vectors, and therefore have a final lateral momentum component given by: $\vec{k}_{\mbox{scatt}}\cdot \hat{x}= K + N k_a$. The $\hat{y}$ component of the wavevector is fixed through the momentum dispersion of a plane wave in a homogeneous medium: $k^2=k_x^2+k_y^2$, so that $k_y=\sqrt{k^2-k_x^2}$. For a certain diffraction order to be propagating, it is required that $k_y\in\Re$ and thus
\begin{equation}
    \left|\frac{K}{k} +N \frac{\lambda}{n a}\right|\leq 1
\end{equation}
In our case, the laser impinges on the grating at normal incidence such that $K=0$. Therefore, in the spectral range of interest ($\lambda\in[500, 600]$ nm), only the zeroth and first order diffraction ($N=0,\pm1$) are relevant. The first diffraction order becomes propagating for $\lambda\leq n a$, which in glass and oil evaluates respectively to $\lambda_{\mbox{glass}}\leq 531$ nm and $\lambda_{\mbox{oil}}\leq 582$ nm. Consequently, within the wavelength region ($\lambda\in[531,582]$ nm) where the first diffraction order is propagating in the oil but not in the glass substrates, guided modes exist in the oil layer. These modes can be interpreted as the diffraction orders radiating in the oil layer and contained by total internal reflection at the oil-glass interfaces.

In figure~\ref{fig:S10} we show  theoretical and experimental transmittance spectra of our sample for different heights of the oil layer. Vertical dashed lines indicate the wavelengths at which the different diffraction orders become propagating, and around which grating resonances can be expected. The vertical green line indicates the incident laser wavelength, which is close a grating resonance as mentioned in the main text. For an oil layer of height $H$, guided mode resonances occur whenever $k_y H=N\pi$. By inserting the previous value for the $\hat{y}$ component of the guided modes wavevector, we see that
\begin{equation*}
    \frac{H}{a}\sqrt{1-\left(\frac{\lambda}{n_{\mbox{oil}}a}\right)^2} = N \frac{\lambda}{n_{\mbox{oil}}a},
\end{equation*}
with $N$ a natural number. Labeling the wavelength of the $N$-th resonance as $\lambda_N$ and taking the limit $\lambda\rightarrow n_{\mbox{oil}}a$, it follows  that
\begin{equation}
    \frac{\lambda_{N+1}-\lambda_{N} }{\lambda_{N+1}+\lambda_{N}}\propto \frac{a}{H}.
\end{equation}
The above expression shows that the spectral density of guided modes increases with the height of the nonlinear domain. Indeed, measurements and simulations in fig.~\ref{fig:S10} show that the density  of guided mode resonances increases with the oil layer height. For the tallest oil layer, the resonances are so sharp that they cannot be properly resolved by our spectrometer.

\subsection{Finite size simulations, observation of diffraction orders in the reflection image} \label{sec:S_dots_linear_simms}
To determine the origin of the bright features enclosed by dashed white circles in Figs.~2(B-D) of the main text, we performed full-wave numerical simulations of our experimental system in COMSOL Multiphysics. The simulation domain consists of a glass-oil-glass waveguide ($n_{oil}=1.59$, $n_{g}=1.45$), with an oil height of 30 $\mu$m, and semi-infinite glass domains of 2 $\mu$m  terminated by perfectly matched layers. The waveguide is 800 $\mu$m wide, and the system is terminated laterally by perfectly matched layers too. On the bottom oil-glass interface stands an aluminium grating of period 366 nm. The individual wires of the grating are rectangular, 100 nm wide and 70 nm tall. To simulate the incident laser beam, we set up a monochromatic Gaussian beam incident from the top with a 10 $\mu$m waist (similar to the experimental setup) and  $\lambda=532$nm wavelength as in experiments.

\begin{figure*}[h]
\centering
\includegraphics[width=\textwidth]{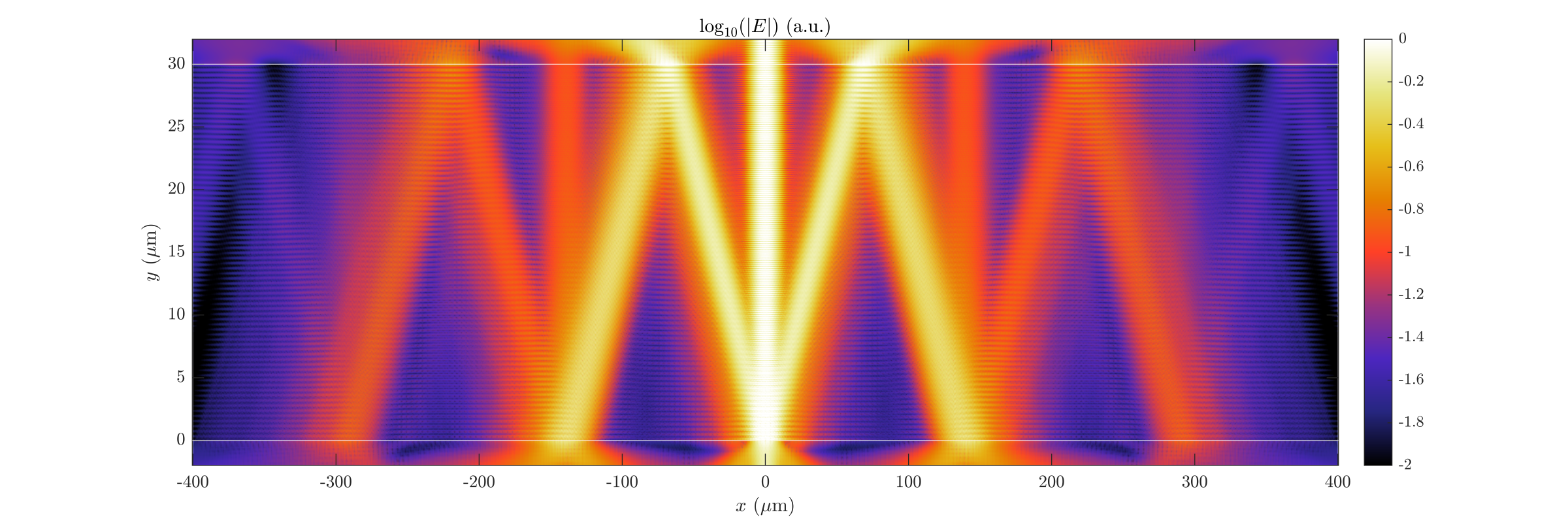}
\caption{\textbf{Simulated electric field profile in our sample illuminated by a Gaussian beam.} The simulation domain consists of a waveguide composed of a two semi-infinite glass layers encapsulating a 30 micron oil layer. An aluminum grating stands on the bottom oil-glass interface, with geometric parameters according to experimental samples. The excitation wavelength is 532 nm, and the Gaussian beam has a $10$ $\mu$m waist.}
\label{fig:S3}
\end{figure*}

Figure~\ref{fig:S3} shows the electric field amplitude in logarithmic scale over the entire simulation domain. As the incident beam enters the oil layer and illuminates the array, the zeroth and $\pm1$ diffraction orders  are excited. Beam replicas are formed along them. The $\pm1$ diffraction orders propagate upwards until they reach the top oil-glass interface. Since the first diffraction order is propagating in the oil but not in the glass, the diffracted waves are contained within the oil layer by total internal reflection. This gives rise to guided modes. Since our parameter range is in the vicinity of the transition of the first diffraction order from evanescent to propagating in glass ($\lambda \leq 530$ nm), diffracted waves are weakly confined and have a long evanescent tail into the glass. As these diffracted waves are reflected into the oil layer and impinge on the grating, another diffraction takes place. This is evidenced by the presence of vertical beams with zero lateral momentum. These vertical beams meet the top oil-glass interface at twice the distance traveled by the first diffraction order in the oil. This simulation shows that when an image is formed on the top oil-glass interface, as in our experiments, the highlighted dots in Figs.~2(B-D) of the main text are replicas of the incident beam formed along the $\pm1$ diffraction orders. Their intensity is therefore directly related to the diffracted intensity by the grating. Note that these dots can be observed due to the finite width of the excitation beam.

\clearpage

\section{Bistability range and thermal relaxation time}
\label{sec:S1_th_relax_overshoot}

\begin{figure}[h]
\centering
\includegraphics[width=\textwidth]{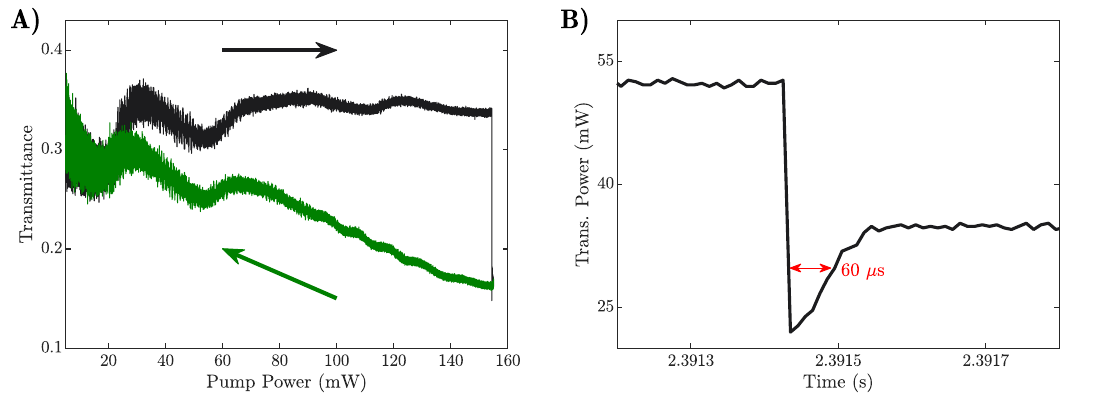}
\caption{\textbf{Bistability range and thermal relaxation time in experiments.} A) Transmittance of the oil-coated grating versus the incident power. This is the same data as in Fig. 2(A) of the main text. Black and green curves correspond to increasing and decreasing power scans, respectively. Their difference evidences a wide bistability range and optical hysteresis. B) Close up view of the transmittance around the time the system switches to a low transmission state. The switch is followed by an undershoot, due to the finite thermal relaxation time of the oil. The half-width at half-maximum of the undershoot,  60 $\mu$s, is indicative of the thermal relaxation time which is also the memory time of our system. }
\label{fig:S8}
\end{figure}
\clearpage

\section{Sketch of the measurement setup}
\begin{figure*}[h]
\centering
\includegraphics[width=\textwidth]{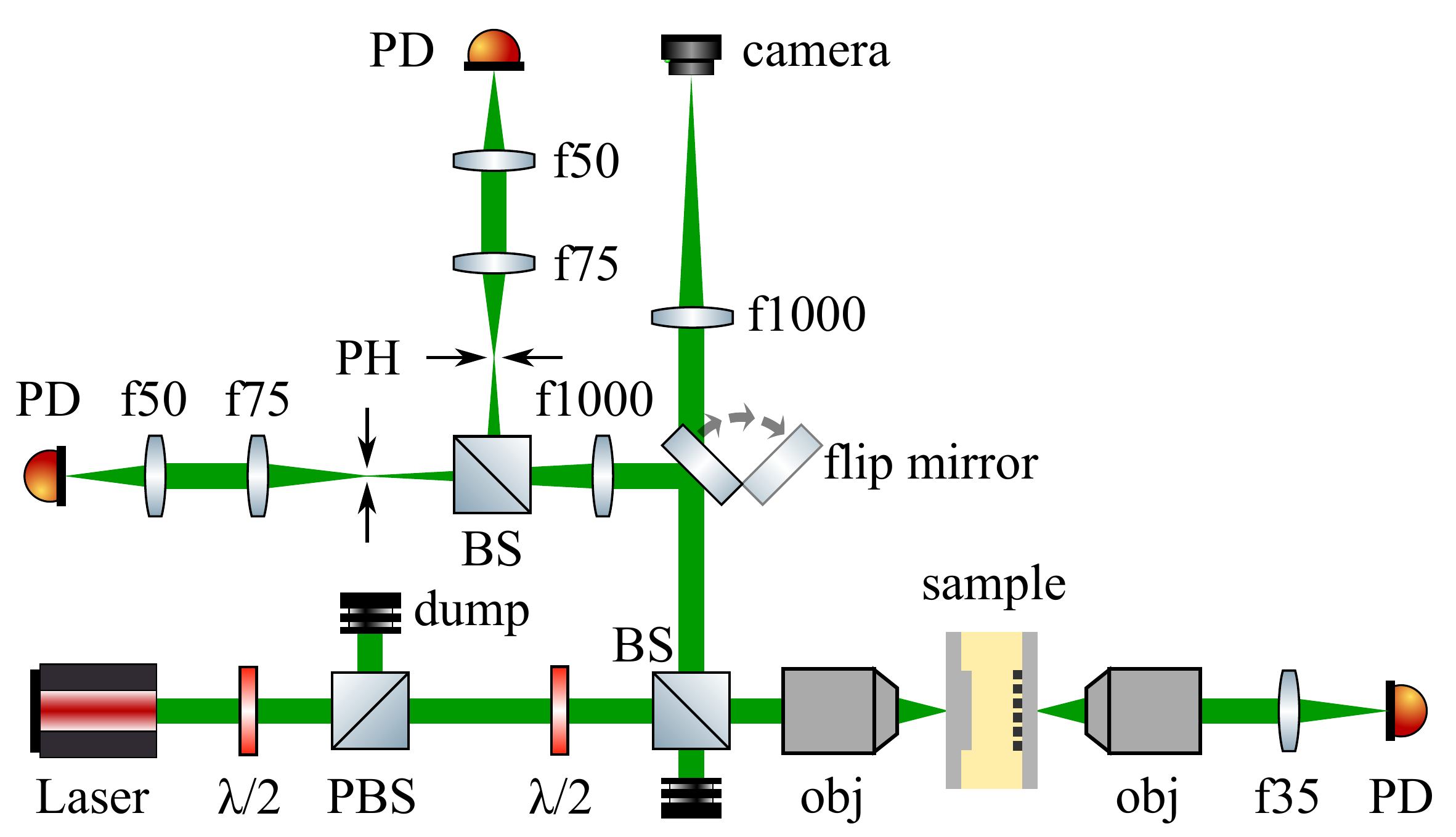}
\caption{\textbf{Setup for the measurements in Fig. 2 and Fig. 3 of the main text.} PBS is polarizing beam splitter; BS is beam splitter; obj is objective; PH is pinhole; PD is photodetector; $\lambda/2$ is half waveplate. The laser intensity is modulated by the PBS and $\lambda/2$. Another $\lambda/2$ is used to ensure that the input laser is linearly polarized parallel to the grating vector before entering the objective. The laser
transmission is measured by a PD. The flip mirror allows sending the reflected intensity either to a camera or to a two-arm setup with photodetectors.}
\label{fig:S11}
\end{figure*}

Figure ~\ref{fig:S11} illustrates the experimental setup we constructed to measure the nonlinear transmission and diffraction of our oil-coated grating. The intensity of a 532 nm continuous-wave laser was temporally modulated using a polarizing beam splitter and a half waveplate installed on a motorized rotational stage. Using another half waveplate, we ensured the alignment of the incident light polarization with the grating vector. We used 10×  (NA = 0.25) microscope objectives for optical excitation and collection. The laser beam, with a diameter of 4 mm before entering the objective (less than the 10 mm aperture of the objective), was loosely focused onto the grating plane. The transmitted intensity was measured by a photodetector. In the reflection path, a flip mirror was incorporated to direct the reflected intensity into either of two separate paths. One directed the intensity to a camera, facilitating spatial data acquisition at a relatively slow rate of 100 frames per second. The alternative path led to a configuration with two arms, each housing a photodetector with 50 MHz bandwidth. These photodetectors measured the diffracted light upon reflection. To effectively isolate the signal from the bright spots depicted in  Figs. 2(B-D) of the main text, which represent the behavior of the ±1 diffraction order of the system as detailed in section \ref{sec:S_dots_linear_simms}, we introduced pinholes at two intermediate image planes. The area of the reflected intensity pattern transmitted  by these pinholes corresponds to the dashed white circles in Figs. 2(B-D) of the main text.

\clearpage

\section{Limit cycles and spontaneous symmetry breaking in electromagnetic fields}

\begin{figure*}[h]
\centering
\includegraphics[trim=0 25 0 25,clip,width=0.85\textwidth]{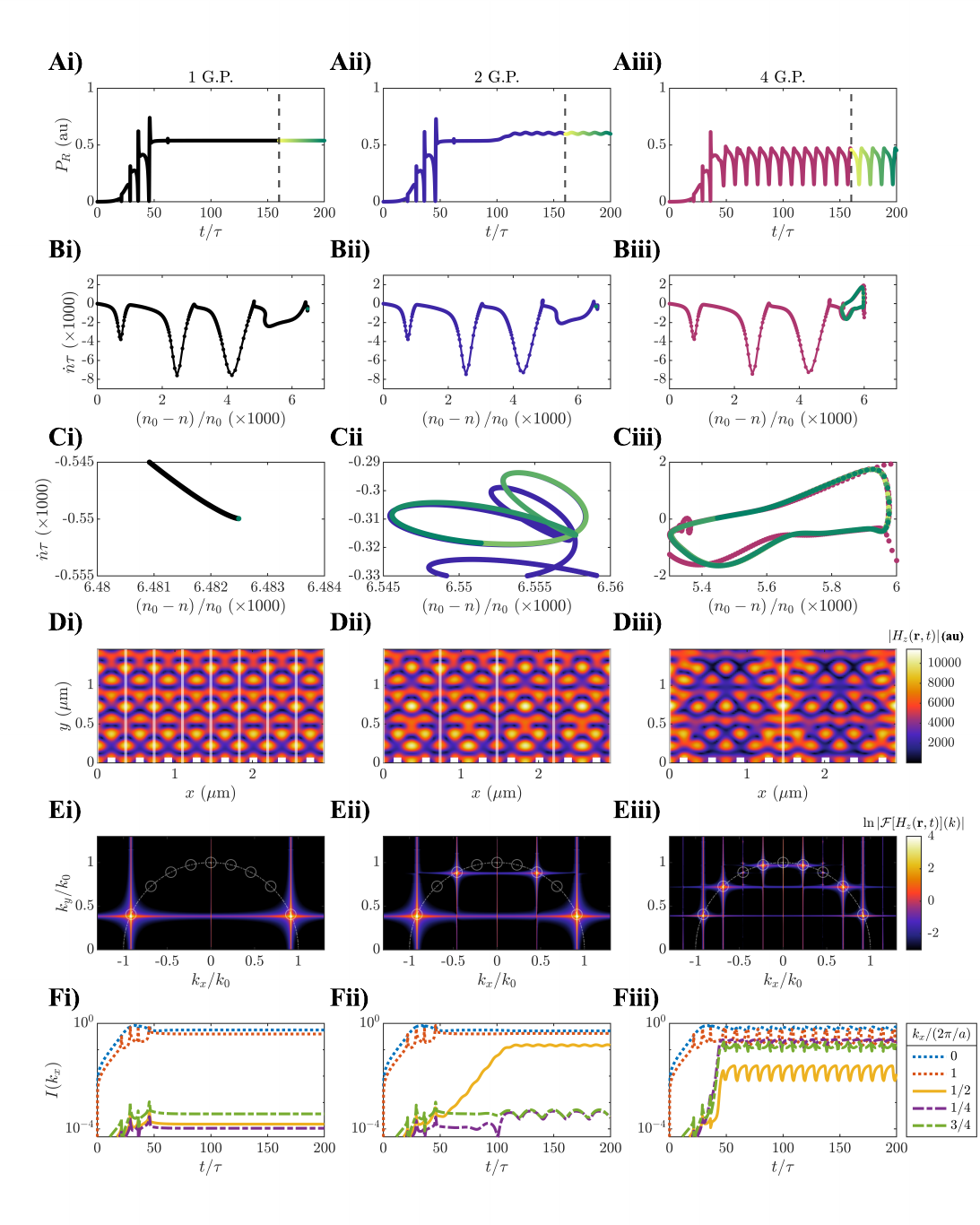}

\caption{\textbf{Limit cycles and breakdown of Bloch's theorem in the electromagnetic fields.} Roman numerals, i-iii (left to right) indicate the number of grating periods (G.P.) in the super-cell. A) Reflected power as a function of time, taken from Fig. 4. B) Phase-plane trajectory of the system (spatial average refractive index vs its time derivative). C) Zoom of the trajectories for $t>160\tau$. Colour of the data-points correspond to the time values indicated in panels (A).
 D) Out of plane component of the magnetic field within the oil layer evaluated at the vertical dashed line shown on panels (A).
 E) Fourier transform of the magnetic fields in (D). F) Integral of the Fourier-amplitude within the small white circles shown in panels (E).}
\label{fig:S14}
\vspace{1cm}
\end{figure*}

Limit cycles correspond to closed periodic orbits in phase space. In systems with few degrees of freedom (DoF), such orbits are easy to visualize on the entire phase space. However, that is not the case of our system. The DoF of our system are the electromagnetic fields at each point in space, which are infinitely many in the continuum limit.  Since we discretize space in our simulations, our number of DoF is actually finite, but nonetheless extremely large. Our challenge is therefore to either isolate the relevant DoF, or follow a coarse graining approach as commonly done in statical physics; we opt for the latter.  In particular, we analyze  the dynamics of our system in the two-dimensional phase space of our coarse grained variables, which are the spatially averaged refractive index of the oil (represented referenced to the linear value as $(n_0-\ev{n})/n_0$) and its time derivative (plotted in adimensional form, $\tau\ev{\dot{n}}=n_0 -\gamma\ev{\left|\mathbf{E} \right|^2} -\ev{n}$). Figures~\ref{fig:S14}(B,C) show the phase space dynamics for the same simulations shown in Fig. 4 of the main manuscript. The colour scale encodes the time evolution as indicated in panels (A). The 1 grating period (G.P.) super-cell displays a steady state solution, which corresponds to a trajectory that collapses onto a single point in phase space. On the other hand, the 2 and 4 G.P. simulations display more complex trajectories which coalesce onto closed orbits in phase space. This demonstrates the existence of limit cycles in our simulations.\\

In figs.~\ref{fig:S14}D(i-iii) we show the amplitude of the out-of-plane magnetic field ($|H_z|$) within the nonlinear regime for 1, 2, and 4 G.P.. The fields are evaluated at $t=160\tau$, indicated by grey vertical lines in Fig. 4A(i-iii) of the main text and in fig. S6A(i-iii). Since the refractive index profiles in Fig. 4B(i-iii)  break Bloch's theorem, it is not surprising that the electromagnetic fields break it too. However, we can extract new insights from the fields by analyzing the Fourier transforms shown in figs.~\ref{fig:S14}E(i-iii). These panels reveal that the electromagnetic solution within the nonlinear regime can be decomposed into a sum of plane-waves that propagate with momentum close to a plane wave in the homogeneous medium. While the wavevector components for 1 G.P. respect Bloch's theorem and only present $k_x=0$ and $\pm2\pi/a$, the solutions for larger super-cells contain momenta forbidden in the linear grating by discrete translation symmetry. This showcases the breakdown of Bloch's theorem in the nonlinear regime. In G.P. 2 and 4, the wavevectors that emerge are limited by the imposed super-cell periodicity. To track their amplitude, $I(k_x)$, in time, we integrate the Fourier transform over circles centered around each of these momentum components, indicated by the small white circles in panels (E). The results are presented in figs.~\ref{fig:S14}F(i-iii), where it can be seen that the Bloch-forbidden wavevector components remain negligible for the 1 G.P., but become significant for larger super-cells. Interestingly, for the 2 G.P. simulation, where the system is slightly above threshold, we observe the exponential growth of the diffraction order $k_x=\pm \pi/a$ for $t\in[50, 100]\tau$. Once this new wavevector component becomes comparable to the linear ones, the reflected power in panel (Aii) begins to oscillate. The fact that the Bloch-forbidden wavevectors and the oscillations in reflection emerge simultaneously evidences the connection between symmetry breaking and limit cycles as observed in our experiments. Figure~\ref{fig:S14}F(i-iii) also evidence that by making the super-cell larger (effectively pushing the system deeper into the non-linear regime), the number of allowed momentum components grows.

\clearpage

\section{Linear Stability Analysis}
Here we provide details and derivations of the linear stability analysis discussed in the main text. Our approach exploits the Born approximation in the solution of Maxwell's Equations for the scattering of light from a metallic grating embedded in a thermo-optical nonlinear medium (oil). It operates in the perturbative limit of weak nonlinear patterning of the oil refractive index.
\subsection{General derivation} \label{sec:S_Born_general_deriv}
The equation of motion for the nonlinear refractive index is
\begin{equation}
\tau \frac{d n(\xx,t)}{dt}+n(\xx,t)=n_0-\gamma\lvert {\bf E}(\xx,\omega,t)\rvert^2
.\end{equation}
In steady-state $\dot{n}(\xx,t)=0$, and the magnitude displayed in Figs.~4B(i-v) of the main text is ${n_0-n(\xx,t)\propto \lvert {\bf E}(\xx,\omega,t)\rvert^2}$. We are interested in the effect of a small index change on the total electric field. In the Born approximation~\cite{Born1999}, if the refractive index is modified by a $small$ amount such that $n\left(\xx,t\right)=n_0+\Delta n(\xx,t)$, the electric field can be written as:
\begin{equation}
    \mathbf{E}\left(\xx,\omega,t\right)=-\frac{2k_0^2}{n_0}\int{d\ \xx^\prime}\ \overline{\overline{G}}\left(\xx-\xx^\prime,\omega\right)\cdot\Delta n(\xx^\prime,t)\mathbf{E}\left(\xx^\prime,\omega,t\right).
\end{equation}
$\overline{\overline{G}}$ is the homogeneous medium Green’s function dyadic for the wave equation, and the integral is performed over the region where the index is modified. Throughout the whole derivation, we use $k_\omega=\omega/c$ to denote the light’s wavevector in vacuum, and $k_0=n_0k_\omega$ to denote the wavevector in the homogeneous medium. For compactness, and since we only consider monochromatic illumination as in experiments,  we omit the frequency dependence from now on.  Born’s approximation tells us that if the index change is small, we can expand the total electric field in powers of the refractive index change. Therefore, the total field reads
\begin{equation}
    \mathbf{E}\left(\xx,t\right)=\sum_{n=0}^{\infty}{\mathbf{E}_n\left(\xx,t\right)}
\end{equation}
\begin{equation}
    \mathbf{E}_{n+1}\left(\xx,t\right)=-\frac{2k_0^2}{n_0}\int{d\ \xx^\prime}\ \overline{\overline{G}}\left(\xx-\xx^\prime\right)\cdot\Delta n(\xx^\prime,t)\mathbf{E}_n\left(\xx^\prime,t\right).
\end{equation}
The first term in the series corresponds to the linear system’s solution $\mathbf{E}_0\left(\xx,\omega\right)$. Since the maximum index modulation is indeed small, we restrict ourselves to the first scattering order, so that the total electric field is approximately given by:
\begin{equation}
\mathbf{E}\left(\xx,t\right)=\mathbf{E}_0\left(\xx\right)-\frac{2k_0^2}{n_0}\int{d\ \xx^\prime}\ \overline{\overline{G}}\left(\xx-\xx^\prime\right)\cdot\Delta n(\xx^\prime,t)\mathbf{E}_0\left(\xx^\prime\right)
\end{equation}
Inserting this expression into the equation of motion for the refractive index yields
\begin{gather}
\begin{split}
&\tau\frac{d\ \Delta n(\xx,t)}{dt}+\Delta n\left(\xx,t\right)=-\gamma\left|\mathbf{E}_0\left(\xx\right)\right|^2+\\
&+2k_\omega^2\gamma\int{d\xx^\prime\left[n_0\mathbf{E}_0^\ast\left(\xx\right)\cdot\overline{\overline{G}}\left(\xx-\xx^\prime\right)\cdot\mathbf{E}_0\left(\xx^\prime\right)+n_0^\ast\mathbf{E}_0^\ast\left(\xx^\prime\right)\cdot{\overline{\overline{G}}}^\ast\left(\xx-\xx^\prime\right)\cdot\mathbf{E}_0\left(\xx\right)\right]\Delta n(\xx^\prime,t)}.
\end{split}
\end{gather}
We now separate $\Delta n(\xx,t)$ into a static and a dynamic part: $\Delta n\left(\xx,t\right)=\delta n_s\left(\xx\right)+\delta n(\xx,t)$. Using the linear field solution, we arrive to an expression for the steady-state index modification
\begin{gather}
\begin{split}
    &\delta n_s\left(\xx\right)=-\gamma\left|\mathbf{E}_0\left(\xx\right)\right|^2+\\
    &+2k_\omega^2\gamma\int{d\xx^\prime\left[n_0\mathbf{E}_0^\ast\left(\xx\right)\cdot\overline{\overline{G}}\left(\xx-\xx^\prime\right)\cdot\mathbf{E}_0\left(\xx^\prime\right)+n_0^\ast\mathbf{E}_0^\ast\left(\xx^\prime\right)\cdot{\overline{\overline{G}}}^\ast\left(\xx-\xx^\prime\right)\cdot\mathbf{E}_0\left(\xx\right)\right]\delta n_s\left(\xx\prime\right)},
    \end{split}
\end{gather}
and for the time evolution of the different perturbations
\begin{gather}
\begin{split}
    &\tau\frac{d\ \delta n\left(\xx,t\right)}{dt}+\delta n\left(\xx,t\right)=\\
    &2k_\omega^2\gamma\int{d\xx^\prime\left[n_0 \mathbf{E}_0^\ast\left(\xx\right)\cdot\overline{\overline{G}}\left(\xx-\xx^\prime\right)\cdot\mathbf{E}_0\left(\xx^\prime\right)+n_0^\ast\mathbf{E}_0^\ast\left(\xx^\prime\right)\cdot{\overline{\overline{G}}}^\ast\left(\xx-\xx^\prime\right)\cdot\mathbf{E}_0\left(\xx\right)\right]\delta n(\xx^\prime,t)}.
\end{split}
\end{gather}
Since we are interested in perturbations that grow over time, we focus on the dynamic part. We first introduce the dyadic Green's function for a 2D system:
\begin{gather}
{\overline{\overline{G}}}_{ij}\left(\xx-\xx^\prime\right)=\left[\delta_{i,j}+\frac{\partial_i\partial_j}{k_0^2}\right]g\left(\xx-\xx^\prime\right)\\
g\left(\xx-\xx^\prime\right)=\frac{i}{4}H_0\left(k_0\left|\xx-\xx^\prime\right|\right)=\int{\frac{d\kk\prime}{4\pi^2}\frac{e^{i\ \kk^\prime\cdot(\xx-\xx^\prime)}}{k^{\prime2}-k_0^2-i\eta}},
\end{gather}
where $g\left(\xx-\xx^\prime\right)$ is the scalar Green’s function of the Helmholtz wave equation in 2D, and $\eta$ is a regularizing factor that tends to $0$. Putting these two together, the complete dyadic can be written as:
\begin{gather}
\overline{\overline{G}}\left(\xx-\xx^\prime\right)=\int{\frac{d\kk\prime}{4\pi^2}\ \overline{\overline{O}}\left(\kk^\prime\right)\ \frac{e^{i \kk^\prime\cdot(\xx-\xx^\prime)}}{k^{\prime2}-k_0^2-i\eta}},
\end{gather}
where
\begin{gather}
\overline{\overline{O}}\left(\kk\right)=\left(\begin{matrix}1-\frac{k_x^2}{k_0^2}&-\frac{k_xk_y}{k_0^2}&0\\-\frac{k_xk_y}{k_0^2}&1-\frac{k_y^2}{k_0^2}&0\\0&0&1\\\end{matrix}\right)
\end{gather}
contains all the tensorial character of the Green's function dyadic. Plugging this back into the equation for the time evolution of the refractive index we get:
\begin{gather}
\begin{split}
\tau\frac{d\ \delta n\left(\xx,t\right)}{dt}+\delta n\left(\xx,t\right)=&\\
2k_\omega^2\gamma\sum_{\alpha,\beta}\iint{d\xx^\prime\frac{d\kk^\prime}{4\pi^2}}\Bigg[& n_0\left(\mathbf{\mathcal{E}}_\alpha^\ast\cdot\ \overline{\overline{O}}\left(\kk^\prime\right)\cdot\mathbf{\mathcal{E}}_\beta\ \frac{e^{i\left(\kk^\prime-\kk_\mathit{\alpha}\right)\cdot\xx}\ e^{i\left(\kk_\mathit{\beta}-\kk^\prime\right)\cdot\xx\prime}}{k^{\prime2}-k_0^2-i\eta}\right)\\
+& n_0^\ast\ \left(\mathbf{\mathcal{E}}_\beta^\ast\cdot{\overline{\overline{O}}}^\ast\left(\kk^\prime\right)\cdot\mathbf{\mathcal{E}}_\alpha\ \frac{e^{i\left(\kk_\mathit{\alpha}-\kk^\prime\right)\cdot\xx}\ e^{i\left(\kk^\prime-\kk_\mathit{\beta}\right)\cdot\xx\prime}}{k^{\prime2}-{(k_0^\ast)}^2+i\eta}\right)\Bigg]\delta n(\xx^\prime,t).
\end{split}
\label{eq:S_Born_b4_space_integrals}
\end{gather}
Here we have introduced that the linear system’s solution can be written as a sum of plane waves as $\mathbf{E}_0\left(\xx\right)=\sum_{\alpha}{\mathbf{\mathcal{E}}_\mathit{\alpha}e^{i\ \kk_\mathit{\alpha}\cdot\xx}}$. To determine the evolution of perturbations of a certain wavevector, we multiply each side by $e^{i\ \kk_\mathit{t}\cdot\xx}/4\pi^2$ and integrate over all space. Writing the perturbation to the refractive index as $\delta n\left(\xx\right)=\int{d\kk\ \delta n\left(\kk\right)\ e^{-i\ \kk\cdot\xx}\ }$, and dropping the explicit time dependence for compactness, we have
\begin{gather}
\begin{split}
&\tau\frac{d\ \delta n\left(\kk_\mathit{t}\right)}{dt}+\delta n\left(\kk_\mathit{t}\right)=\\
&2k_\omega^2\gamma\sum_{\alpha,\beta}{\left[n_0\left(\ \frac{\ \mathbf{\mathcal{E}}_\alpha^\ast\cdot\ \overline{\overline{O}}\left(\kk_\mathit{\alpha}-\kk_\mathit{t}\right)\cdot\mathbf{\mathcal{E}}_\beta}{\left|\kk_\mathit{\alpha}-\kk_\mathit{t}\right|^2-k_0^2-i\eta}\right)+n_0^\ast\ \left(\ \frac{\ \mathbf{\mathcal{E}}_\alpha^\ast\cdot{\overline{\overline{O}}}^\ast\left(\kk_\mathit{\beta}+\kk_\mathit{t}\right)\cdot\mathbf{\mathcal{E}}_\beta}{\left|\kk_\mathit{\beta}+\kk_\mathit{t}\right|^2-\left(k_0^\ast\right)^2+i\eta}\right)\right]\delta n\left(\kk_\mathit{t}+\kk_\beta-\kk_\alpha\right)\ }.
\end{split}
\end{gather}
Here we have assumed that the nonlinear domain extends infinitely. For brevity, we introduce $F_{\alpha\beta}\left(\kk\right)=\frac{\mathbf{\mathcal{E}}_\mathit{\alpha}^\ast}{|E_0|}\cdot\left[\mathbb{I}-{\frac{\mathbf{k}\mathbf{k}}{k_0^2}}\right]\cdot\frac{\mathbf{\mathcal{E}}_\mathit{\beta}}{|E_0|}$, with $|E_0|$ being the amplitude of the incident plane wave. By taking into account a small absorption in the host medium, we can drop the regularizing factor $\eta$ from the Green's function and write the refractive index as $n_0=n_{0,r}+in_{0,i}=n_{0,r}\left(1+i\chi\right)\approx|n_0|(1+i\chi)$, where $\chi\ll1$. Then the equation above reads
\begin{gather}
\begin{split}
  \tau\frac{d\ \delta n\left(\kk_\mathit{t}\right)}{dt}+\delta n\left(\kk_\mathit{t}\right)=&\\
  =\frac{2}{\left|n_0\right|}\frac{\gamma\left|E_0\right|^2}{\chi}\sum_{\alpha,\beta}\Bigg[(1+i\chi)&\left(\ \frac{\chi F_{\alpha\beta}\left(k_\alpha-k_t\right)}{\frac{\left(k_\alpha-k_t\right)^2}{\left(\left|n_0\right|k_\omega\right)^2}-(1-\chi^2+2i\chi)}\right)\\
  +(1-i\chi)&\left(\ \frac{\chi F_{\beta\alpha}^\ast\left(k_\beta+k_t\right)\ }{\frac{\left(k_\beta+k_t\right)^2}{\left(\left|n_0\right|k_\omega\right)^2}-(1-\chi^2-2i\chi)}\right)\Bigg]\delta n\left(\kk_\mathit{t}+\kk_\beta-\kk_\alpha\right)\approx\\
  \approx\frac{2}{\left|n_0\right|}\frac{\gamma\left|E_0\right|^2}{\chi}\sum_{\alpha,\beta}&\Bigg[\frac{\chi F_{\beta\alpha}\left(k_\beta-k_t\right)}{\frac{\left(k_t-k_\beta\right)^2}{\left(\left|n_0\right|k_\omega\right)^2}-2i\chi-1}+\frac{\chi F_{\alpha\beta}^\ast\left(k_t+k_\alpha\right)\ }{\frac{\left(k_t+k_\alpha\right)^2}{\left(\left|n_0\right|k_\omega\right)^2}+2i\chi-1}\Bigg]\delta n\left(\kk_\mathit{t}+\kk_\alpha-\kk_\beta\right)\
  \end{split}
\end{gather}
Finally, we arrive at the equation presented in the main text:
\begin{gather}
    \tau\frac{d\ \delta n\left(\kk\right)}{dt}+\delta n\left(\kk\right)=\frac{2}{\left|n_0\right|}\frac{\gamma\left|E_0\right|^2}{\chi}\sum_{\alpha,\beta}{\mathcal{M}_{\alpha\beta}(\kk)\delta n\left(\kk+\kk_\alpha-\kk_\beta\right)\ },\label{eq:SBorn_index}\\
\mathcal{M}_{\alpha\beta}(\kk)=\frac{\chi F_{\beta\alpha}\left(\kk_\mathit{\beta}-\kk\right)}{\frac{\left|\kk-\kk_\mathit{\beta}\right|^2}{k_0^2}-1-2i\chi}+\frac{\chi F_{\alpha\beta}^\ast\left(\kk+\kk_\mathit{\alpha}\right)\ }{\frac{\left|\kk+\kk_\mathit{\alpha}\right|^2}{k_0^2}-1+2i\chi}.\label{eq:SBorn_index_couplings}
\end{gather}
Eqs. \ref{eq:SBorn_index} and \ref{eq:SBorn_index_couplings} have several interesting properties. Firstly, Eq. \ref{eq:SBorn_index} is non-local in k-space, since it connects every k-point, $\kk$, with all those given by the difference of two wavevectors present in the linear solution, $\kk_\alpha$, $\kk_\beta$. For this reason, standard numerical toolboxes designed to solve differential equations are of little help and we have to design our own numerical approach. It is important to note that in our particular case, since all the wavevectors present in the linear system correspond to diffraction orders of a grating, then only wavevectors with $k_x$ that differ by a whole number of reciprocal lattice vectors will couple to each other. Secondly, looking at the structure of the couplings in Eq. \ref{eq:SBorn_index_couplings}, one can see that the couplings diverge for wavevectors that are a distance $k_0$ from wavevectors present in the linear solution ($\kk_\alpha$ and $\kk_\beta$). This suggests that those wavevectors will be the most relevant ones in the  nonlinear regime. Indeed, when we draw these white circles over the fourier transform of the refractive index maps on Figure~4C(i-v), we see a good agreement on the location of the main contributions to the index profile.\\

To study the dynamics predicted by Eq.\ref{eq:SBorn_index}, as presented on the main text, we look for the eigenfunctions $\delta \tilde{n}_\lambda({\bf k},t)$ that satisfy
\begin{gather}
\sum_{\alpha,\beta}\mathcal{M}_{\alpha\beta}(\kk)\delta \tilde{n}_\lambda(\kk+\kk_\alpha-\kk_\beta,t)=\lambda \delta \tilde{n}_\lambda(\kk,t).\label{eq:Seig_prob_born}
\end{gather}
By solving Eq. \ref{eq:SBorn_index}, we can now determine the dynamics of the different eigenfunctions:
\begin{gather}
\delta \tilde{n}_\lambda(\kk,t)=\delta \tilde{n}_\lambda(\kk,0)\exp\left[ \left(-1+\frac{2}{\left|n_0\right|}\frac{\gamma\left|E_0\right|^2}{\chi}\lambda \right)\frac{t}{\tau}\right].
\end{gather}
For eigenfunctions corresponding to eigenvalues with positive real part, it is possible to define a critical input field for which exponential growth of the perturbation is guaranteed: $\gamma\left|E_{\mbox{c}}\right|^2 = |n_0|\chi/2 \mbox{Re}\left(\lambda\right)$. The eigenfunction with the largest $\mbox{Re}(\lambda)$ will have the lowest associated threshold, and as such will be the dominant perturbation when the system is driven into the nonlinear regime. If the eigenvalues are complex, then the amplitude of these perturbations will oscillate in time with a frequency given by $\omega_\lambda=2\gamma \left|E_0\right|^2 \mbox{Im}\left(\lambda\right) / \left|n_0\right| \chi \tau$. This shows that the memory time of the system establishes an order of magnitude for the timescale of the dynamics in the nonlinear system. However, more generally, the dynamics are also influenced by how far above the critical field the system is driven.




\subsection{Numerical discretization and diagonalization}
\begin{figure}[t]
\centering
\includegraphics[width=0.5\textwidth]{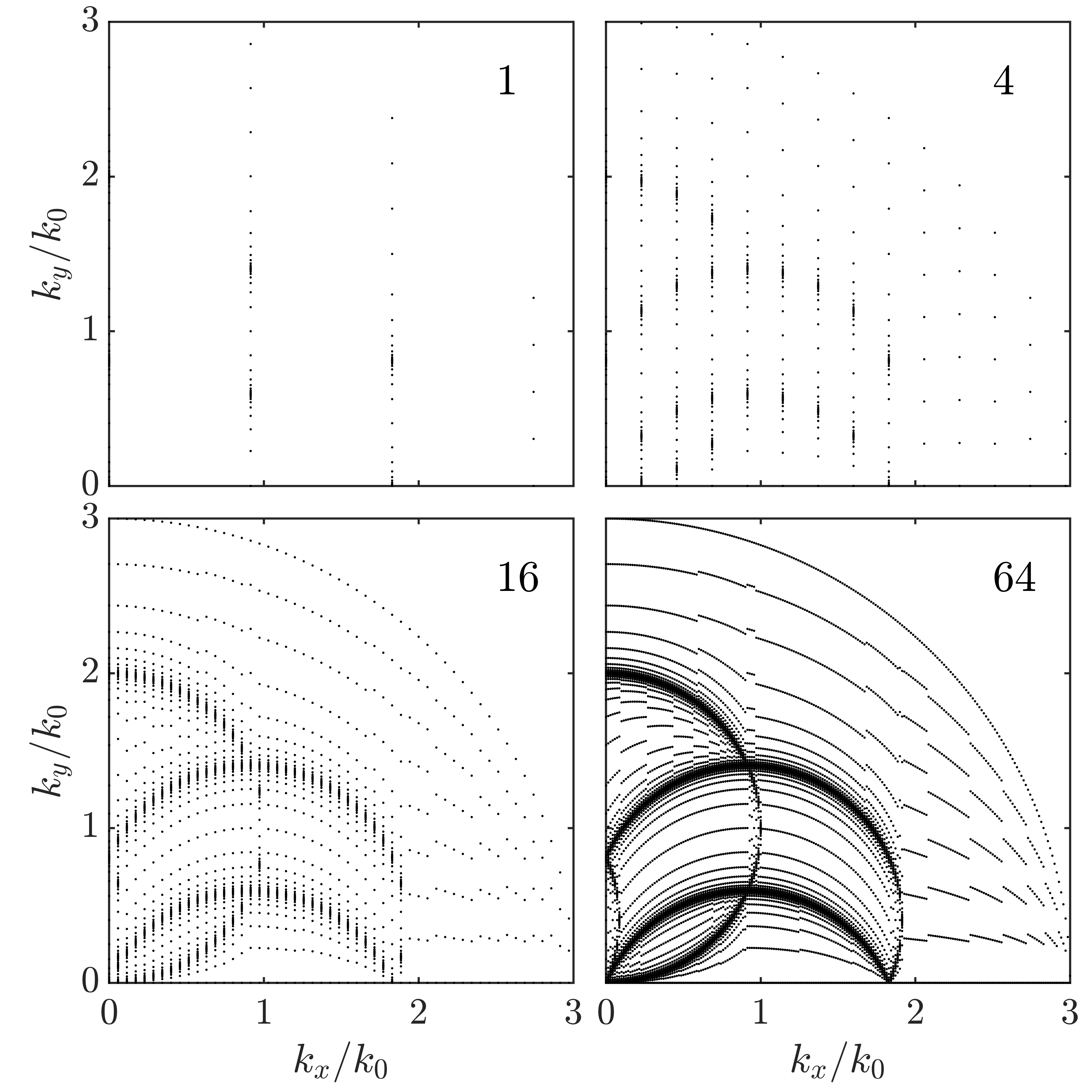}
\caption{\textbf{k-space sampling for the numerical diagonalization.} $k_x$ values discretized according to spacing $\Delta k_x=2\pi/Na$. Different panels correspond to different choices of $N$.}
\label{fig:S1}
\end{figure}

To numerically solve the eigenvalue problem in Eq. \ref{eq:Seig_prob_born}, we first discretize k-space and then write the eigenfunctions as a vector containing the value of the eigenfunction on each of the discrete k-space points. This turns the continuous eigenfunction problem in an effective eigenvector problem. High quality eigenfunctions (eigenvectors) should be obtained under dense enough meshing of k-space. Luckily, we can use several properties of the structure of the eigenvalue problem above to our advantage:
\begin{itemize}
    \item Since the coupling coefficients are divergent at the circles of radius $k_0$ around the wavevectors given in the linear solution, we can make the mesh denser around these circles and more sparse away from them.
    \item Since the refractive index modification is a real quantity, one can show that $\delta n (\kk)=\delta n^*(-\kk)$, and therefore by applying this symmetry to the eigenfunctions  we can restrict the meshing to half of the momentum plane.
    \item Since only wavevectors with $k_x$ that differ by a whole number of reciprocal lattice vectors ($\vec{G}=2\pi/a \hat{x}$) are coupled, then we can discretize $k_x$ values by setting the $k_x$ spacing to be a fraction of the reciprocal lattice wavevector ($\Delta k_x=2\pi/Na$). Doing this is equivalent to setting the overall period of the system to be that of $N$ grating periods.
\end{itemize}
Applying these rules, we generate the k-space samplings shown on fig. \ref{fig:S1}, where we show only the positive quarter plane of momentum space. We limit modulus of momentum values to $3k_0$. Note that the sampling is symmetric with respect to reflections from the $k_y$ axis.
Once the k-space sampling is obtained, the coefficient matrix is built by evaluating Eq. \ref{eq:Seig_prob_born} at every $\kk$ point of our discretization mesh. Every $\kk$ point is coupled to a set of of target wavevectors given by $\kk+\kk_\alpha-\kk_\beta$. For the terms with $\alpha=\beta$, this just gives diagonal contributions to the matrix, and the corresponding $\kk$ values are (obviously) contained exactly within the discretization. On the other hand, for $\alpha\neq \beta$ the target momentum value might not be exactly contained in the discretization. However, we know that at least the $\hat{x}$ component of the target wavevector will be contained in our discretization. Therefore, from our discrete set of k-space sampling, we look for momentum values with matching $k_x$ component to the target momentum, and look for the two closest ones. Then we distribute the coupling between these two according to how close the target momentum was to the sampling points. This may effectively lead to introducing some artificial width into the eigenfunctions. However, since the relevant regions of momentum space are densely meshed, negligible error is introduced when following this approach.\\

\begin{figure}[t]
\centering
\includegraphics[width=0.7\textwidth]{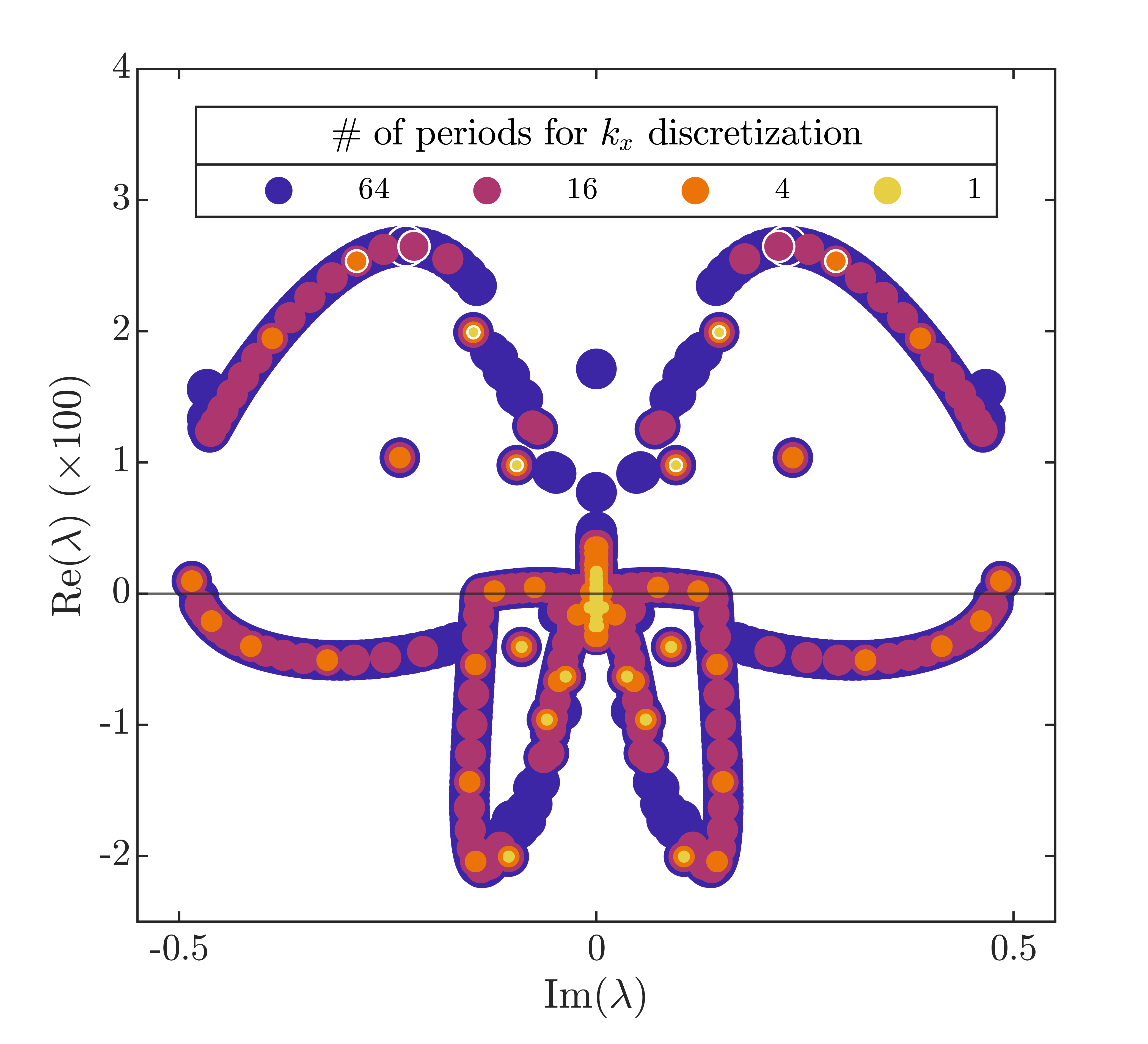}
\caption{\textbf{Eigenvalues calculated for different imposed periodicities.} The 4 eigenvalues with the largest real part are highlighted with a white edge for each imposed periodicity.}
\label{fig:S2}
\end{figure}

After building the coefficient matrix, we use MATLAB's pre-built routine for numerical diagonalization and we extract eigenvalues and eigenfunctions for the different discretizations. On fig.~\ref{fig:S2} we show the numerically obtained eigenvalues for the discretizations shown on fig.~\ref{fig:S1}. Note that all eigenvalues appear with the complex conjugate counterpart. This stems from the refractive index being a real magnitude, and these complex conjugated eigenvalues will appear at the same time when the appropriate critical field is crossed. We can observe how the eigenvalues that belong to discretizations with ever larger $N$ seem to coalesce along continuous curves, giving us confidence that in the limit of $N\rightarrow \infty$ the behavior will be similar as the one shown here. The four governing eigenvalues for each discretization (those with the largest real part) are highlighted by white edges. We can observe that the maximum ${\rm Re}\{\lambda\}$ increases with $N$, which in turn implies that $E_c$ diminishes with increasing supercell size, in agreement with the numerical simulations in Fig.~4. In all cases these eigenvalues have a non-zero imaginary part, which implies that every imposed periodicity experiences a Hopf bifurcation in which the temporal symmetry of the system will be spontaneously broken. It also turns out that these dominant eigenvalues are doubly degenerate in all cases except for $N=1$. This double degeneracy is an indication that whenever the system crosses the given nonlinear threshold, it may evolve according to the two eigenfunctions associated with the degenerate eigenvalues. As we show in Figure~5 of the main text, these eigenfunctions are related by reflection symmetry, and therefore lead to spatial spontaneous symmetry breaking. Also, the resulting eigenfunctions are sharply peaked around the momentum values located along the aforementioned circles in k-space, which justifies our momentum-space sampling strategy.\\

\begin{figure}[t]
\centering
\includegraphics[width=0.7\textwidth]{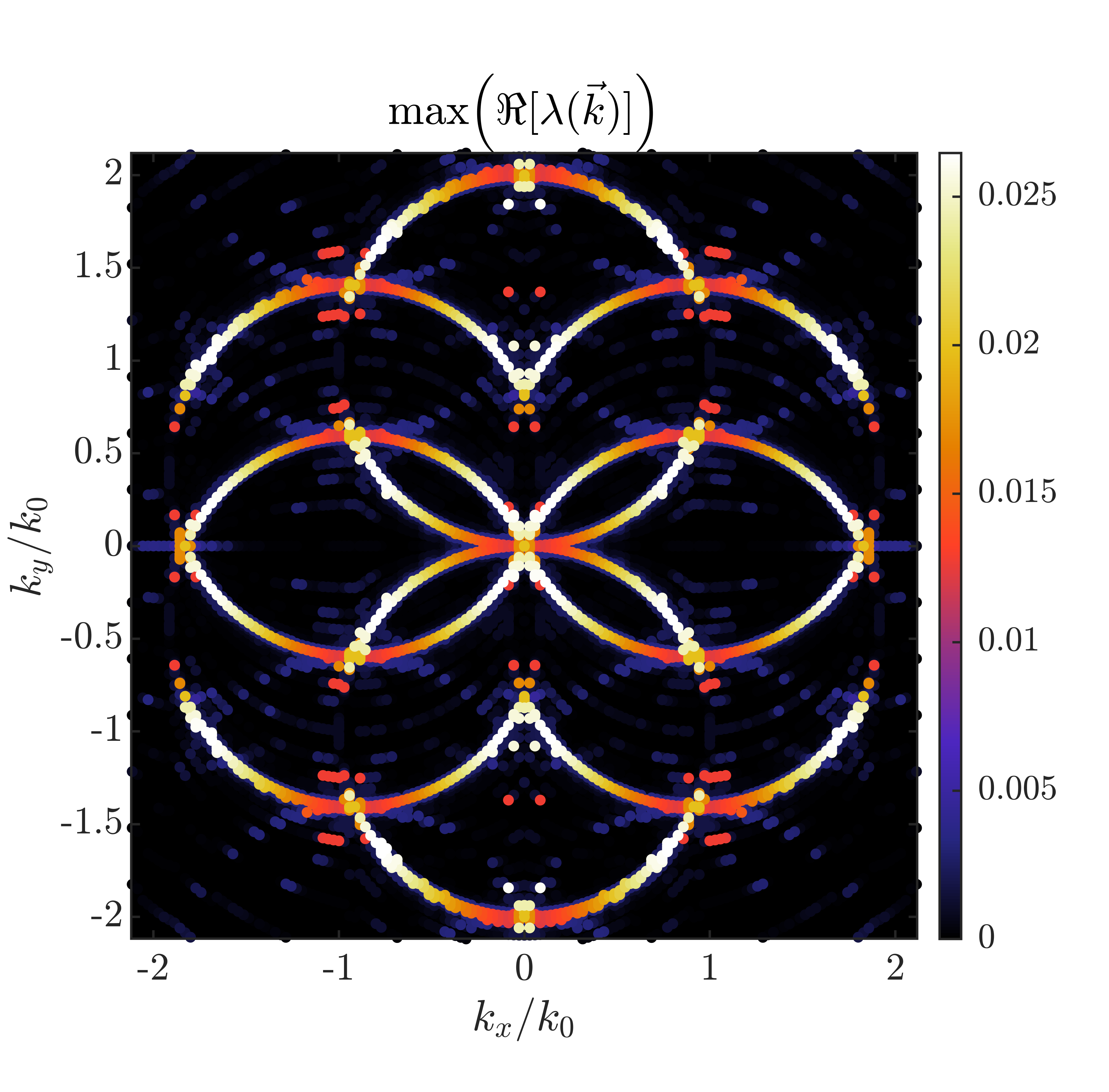}
\caption{\textbf{Maximum positive eigenvalue associated with every wavevector in our discretization.} Discretization mesh corresponding to the case of a 32 grating period supercell.}
\label{fig:S13}
\end{figure}

Until now we have focused on the dominant perturbations. Now we broaden our view to take a look at perturbations with an eigenvalue near the dominant one. On fig. \ref{fig:S13} we show the maximum positive eigenvalue that involves every wavevector component of our discretization i.e. at what critical power one expects to see a perturbation with said wavevector appear. This diagonalization corresponds to a supercell of 32 grating periods. Interestingly, the momentum values associated with the dominant eigenvalue are inmediately surrounded by momenta with very close eigenvalue. This implies that as the system is driven into the nonlinear regime, wavevectors that are very close in momentum space will tend to appear, creating a cellular pattern \cite{Hoyle2006}. This behavior is confirmed by the full numerical simulations. Furthermore, the fact the largest eigenvalues of the system lie along the aforementioned circles in k-space means that as the system is driven further into the nonlinear regime, and reaches chaos, all these momenta will be involved in the dynamics as shown in Figure~4C(v) of the main text.

\clearpage
\newpage


\end{document}